\documentclass{article}
\usepackage{epsfig,amsmath,amsfonts,graphicx}

\begin{document}
\title{Delayed Random Relays}

\author{Koki Sugishita and Toru Ohira\\
Graduate School of Mathematics, Nagoya University, Japan}
\maketitle

\begin{abstract}
 We present here a system with collection of random walks relaying a signal in one dimension in the presence of delays. We are interested in the time for a signal to travel from one end (start) to the other end (finish) of the lined group of random walkers. The delay is introduced at the point when the signal is transferred from each walker to the next one. It is found that there is an optimal number of walkers for the signal to travel fastest when delays are present. We discuss implications of this model and associated behaviors to physical and biological systems.
\end{abstract}

\section{Introduction}

A Random walk is one of the simplest stochastic processes. Yet, it produces remarkably rich behaviors. Also, it has wide applications and has been studied in mathematics, physics, chemistry and so on\cite{bailey,berg,vankampen,weiss}.
 A collection of random walkers has also gained attentions, investigated under the names of multiple random walks, branching random walks, diffusion limited aggregations and so on\cite{aldous,cooper,shi,witten}. 

We will propose and investigate a model with a collection of random walks. Random walkers which are lined and move in one dimension passes a ``message'' from one to the next. The message holder can pass it to the next one when they come in contact.
We investigate behaviors during this relaying of the message from the starting walker to the final one. We further introduce delays in each transfer of the message, and calling our model as 
``Delayed Random Relay''(DRR). In this model, each walker is required to hold the message for a certain time period (delay) before it can pass to the next walker. We will investigate the effect of this inclusion of the delay on the total relaying time.

Delays in dynamics have been studied, particularly with feedback control systems\cite{glassmackey1988,stepan1989}. 
With delays, even a simple first order ordinary differential equation can show rather complex behaviors with delays. They have been studied under the name of ``Delay Differential Equations''. By further introducing stochastic elements, ``Stochastic Delay Differential Equations'' \cite{kuckler1,frankbeek2001} or ``Delayed Random Walks''\cite{ohiramilton,ohirayamane} are proposed and investigated. 
We can consider DRR we discuss here as one example of models incorporating stochasticity and delay.

First, we describe our model of relaying by random walkers and incorporation of delays to formulate the DRR. We investigate the model with a particular focus on how delays affect the total relaying time of the message. 
It will be shown  by computer simulations that a non-monotonic behavior of the total relaying time as a function of the 
number of random walkers emerge with the introduction of delays. 
We will discuss its relation to ``Delayed Stochastic Resonance''\cite{ohirasato}, and possible applications.

\section{Relaying by Random Walkers}

 Let us start with a description of our model without delays. (Figure 1).
We consider a one-dimensional line with a periodic boundary condition (a circle).
On the line, there are $N$ discrete sites on which random walkers hop.
 We place $n$ simple symmetric random walkers on these sites. Each random walker, at each unit time, takes a unit step  either to his right or left with the equal probabilities of $1/2$.
\begin{figure}[h]
\begin{center}
\includegraphics[width=0.9\columnwidth]{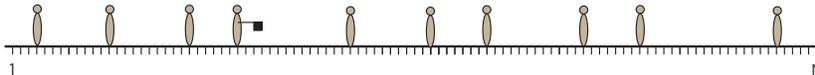}
\end{center}
\caption{A schematic view of relaying by random walkers. We show the case with 10 walkers. The forth walker 
from the left is holding a message, which moves to the right.}
\label{Fig1}
\end{figure}

A message is relayed in one direction by this group of random walkers. When the message holder moves to an adjacent position, i.e., in contact, to the next walker,  the message is passed on. We measure the time, $T$, for the message to travel from the starting random walker to the last walker in the line. 

We performed computer simulations of this model with the parameters of $N=100$ and varying the number $n$ of walkers in the relay. 
For each $n$, we repeated the simulations for 10000 trials and obtained 
the average values of this total relaying time $T$ which is plotted as a function of $n$ (Figure 2).
We observe a monotonic decrease of $T$, meaning that the more walkers we have in the relay, the faster the
message is conveyed.

\begin{figure}[h]
\begin{center}
\includegraphics[width=0.8\columnwidth]{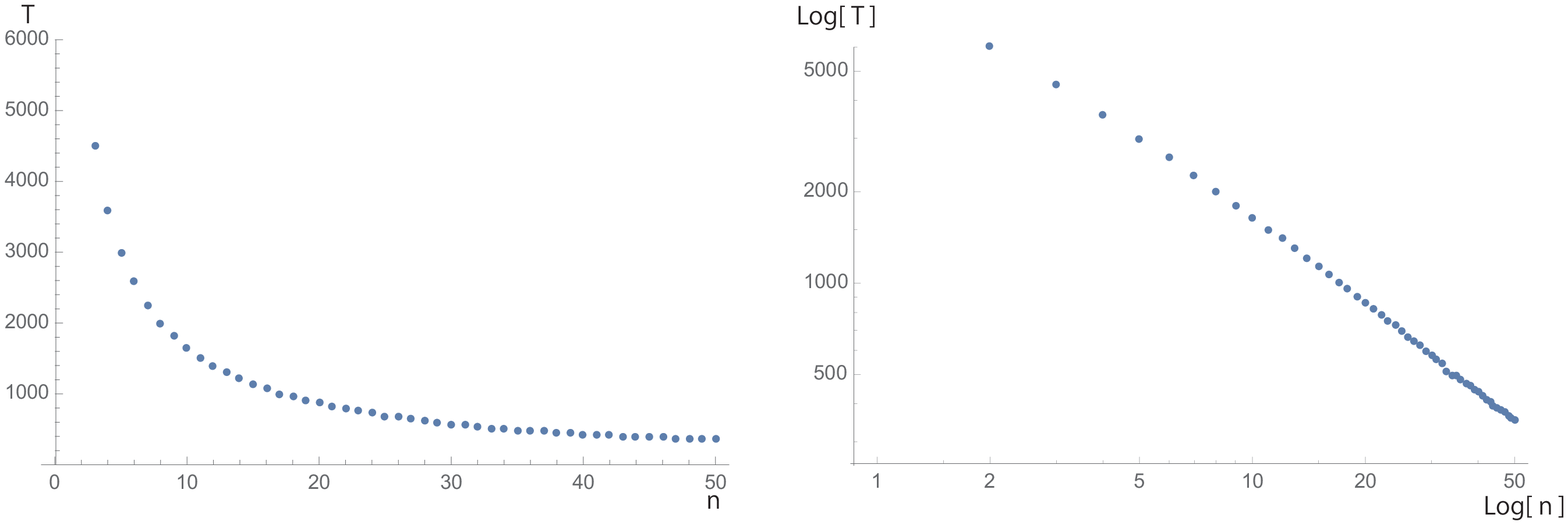}
\end{center}
\caption{Total relaying time $T$ as a function of the number of walkers $n$. The right graph is a Log-Log plot of the same data.}
\label{Fig2}
\end{figure}

\section{Delayed Random Relay}

Now we extend the basic model in the previous section to include a delay in message passing. When the message is received 
from previous walker, it is not immediately transferable to the next walker. There is a time interval, which we call delay $d$, 
for the message to become transferable. One can view this as a requirement for each walker to hold the message for at least a certain time interval, or as a maturing time for the message becoming ready to move on.

Hence, even though a message holding random walker comes in contact with the next walker,  he may not be able to
pass the message. Only after the other condition that the delay time has elapsed is also satisfied, the message is passed on.

With this extended model, we also measured the total relaying time $T$ as a function of $n$. A representative result is shown in Figure 3. Notable feature is that there exists an optimal number of walkers for the fastest relaying of the message.

\begin{figure}[h]
\begin{center}
\includegraphics[width=0.8\columnwidth]{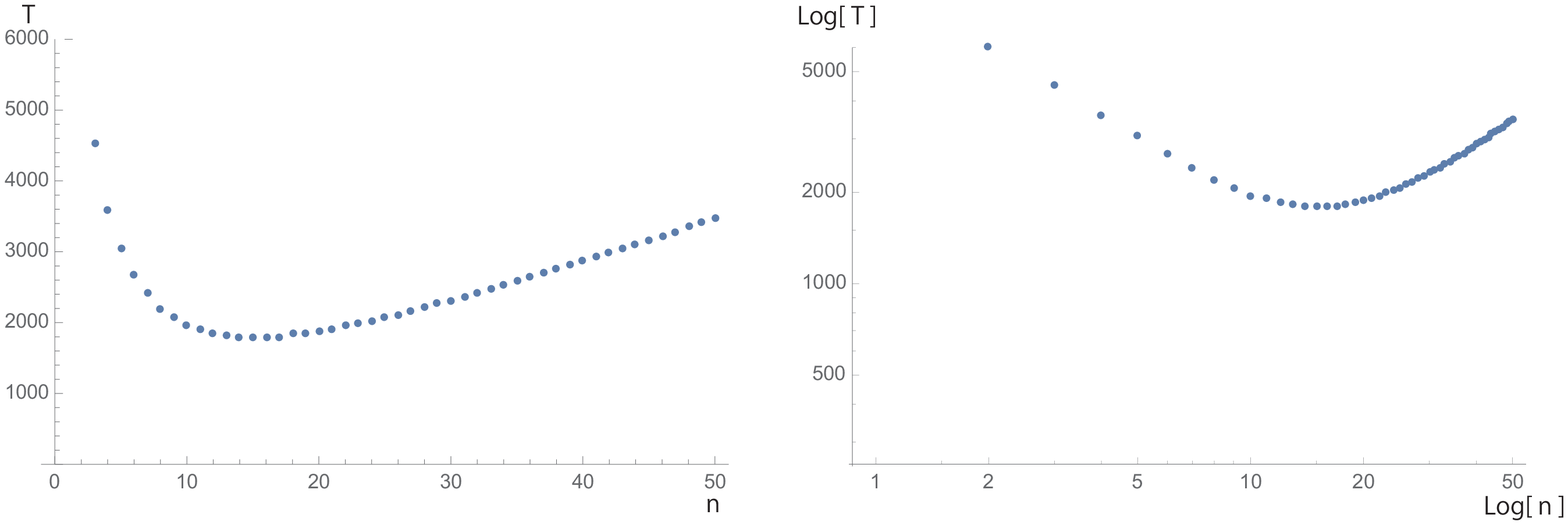}
\end{center}
\caption{Total relaying time $T$ as a function of the number of walkers $n$ when there is a delay of $d=64$. The right graph is a Log-Log plot of the same data.}
\label{Fig3}
\end{figure}

\section{Qualitative Analysis}

The existence of the optimal number of walkers in the relay can qualitatively understood as follows. With a large number of walkers, the average distance between the walkers is small. Thus, the average time of their contacts is short.
If the delay time is longer than this average contact time, the total relaying time is prolonged. 
When the number of walkers is small, the delay time is more likely shorter than the average contact time. 
In this case, the total relaying time is not affected. The optimal number of walkers to have the shortest relaying time
is in between. 

We can roughly estimate this optimal number of walkers. The average contact time between the two walkers is roughly the square of the average distance $N/n$ between the walker. If we set $n^*$ as the optimal number of the walkers, 
the following relation is obtained.

$$ d \approx ({N \over {n^*}})^2 $$

or

$$ n^* \approx {N \over {\sqrt{d}}} $$

In Figure 3, we see the minimum relaying time occurs at around $n \approx 15$. The estimation from
the above equation is $n^* \approx 13$. Our preliminary simulation results indicate that this rough estimation is 
justified to a certain degree with other parameter values.

\section{Discussion}

We have presented a model of relaying a message by random walkers in the presence of delay.
The total relaying time decreases monotonically with increasing number of walkers when there is no delay in passing the message between the walkers. With delays, however, this monotonicity is broken, and there is an optimal number of
walkers for the fastest relay passing of the message. Even though, more quantitative analysis is needed, we can
qualitatively estimate this optimal number as a function of the delay.

There have been studies of ``Stochastic Resonance'', which is a resonant-like behavior with an optimally tuned level of noise and oscillatory signals\cite{benzi,wiesenfeld-moss95,bulsara96,gammaitoni98}.  We can obtain a similar behavior when we replace oscillatory signals by oscillation due to delay. This is called  ``Delayed Stochastic Resonance'', and has been studied both theoretically and experimentally\cite{ohirasato,pikovsky,misono}. One may view the DDR as a collective version of Delayed Stochastic Resonance models.

If there is a process, where the transfer of energy (heat, light, and so on) is much longer than known mechanisms, there is a possibility of delays in the micro-transfer among the basic constitutes. For an example, the sun generates its thermal energy at its core, but it takes unusually long time for the energy to go through the radiative zone to reach the sun's surface. The estimation of this time ranges from 170,000 to 50,000,000 years. The detailed mechanism is still unknown. Though quite rough, 
we may be able to estimate the delay in micro-transfer among the particles involved, if we can estimate the number of particles, 
or the density of various parts of the sun.

Another area of application is an epidemic spreading with latent periods. By observing spreading speed and number of infected in a large scale, one may infer the latent periods. Qualitatively similar behavior may arise in rumor spreadings or passing down of folklore stories.

\section{Acknowledgments} 
T. O. would like to thank Profs. Y. Nakamura, H. Ohira, and K. Todayama of Nagoya University for discussions and for their comments.
This work was in part supported by research funds from Ohagi Hospital (Hashimoto, Wakayama, Japan), Grant-in-Aid for Scientific Research from Japan Society for the Promotion of Science No. 16H01175, and No. 16H03360.

\end{document}